%% file: EXOT-2012-13.tex
\documentclass[11pt,a4paper]{article}
\usepackage{jheppub}
\usepackage{atlasphysics}
\usepackage{subfigure}
\usepackage{mathrsfs}
\usepackage{url}
\usepackage{authblk}
\usepackage{multirow}
\usepackage{lineno}

\def\ee{\ensuremath{e^{\pm}e^{\pm}}}
\def\emu{\ensuremath{e^{\pm}\mu^{\pm}}}
\def\mm{\ensuremath{\mu^{\pm}\mu^{\pm}}}
\def\ll{\ensuremath{\ell^{\pm}\ell^{\pm}}}

\def\ptone{\ensuremath{p_{\mathrm{T}1}}}
\def\pttwo{\ensuremath{p_{\mathrm{T}2}}}

\usepackage{preprintcover}
\PreprintCoverPaperTitle{Search for anomalous production of prompt like-sign lepton pairs at $\sqrt{s} = 7$~TeV with the ATLAS detector}
\PreprintIdNumber{CERN-PH-EP-2012-249}
\PreprintCoverAbstract{An inclusive search for anomalous production of
  two prompt, isolated leptons with the same electric charge is
  presented.  The search is performed in a data sample corresponding
  to 4.7 \ifb\ of integrated luminosity collected in 2011 at $\sqrt{s}=7$~TeV 
  with the ATLAS detector at the LHC. Pairs of leptons ($\ee$, $\emu$, and $\mm$) with large transverse momentum are selected, and 
  the dilepton invariant mass distribution 
  is examined for any deviation from the Standard Model expectation.
  No excess is found, and upper limits on the production cross section of like-sign lepton pairs
  from physics processes beyond the Standard Model are placed as a
  function of the dilepton invariant mass within a fiducial region close to the experimental 
  selection criteria. The 95\% confidence level upper limits on the cross section of 
  anomalous $\ee$, $\emu$, or $\mm$ production range between 1.7~fb and 64~fb depending 
  on the dilepton mass and flavour combination. }
\PreprintJournalName{Journal of High Energy Physics}

\title{Search for anomalous production of prompt like-sign lepton pairs at $\sqrt{s} = 7$~TeV with the ATLAS detector}

\usepackage{authblk}

\collaboration{ATLAS Collaboration}

\abstract{An inclusive search for anomalous production of
  two prompt, isolated leptons with the same electric charge is
  presented.  The search is performed in a data sample corresponding
  to 4.7 \ifb\ of integrated luminosity collected in 2011 at $\sqrt{s}=7$~TeV 
  with the ATLAS detector at the LHC. Pairs of leptons ($\ee$, $\emu$, and $\mm$) with large transverse momentum are selected, and 
  the dilepton invariant mass distribution 
  is examined for any deviation from the Standard Model expectation.
  No excess is found, and upper limits on the production cross section of like-sign lepton pairs
  from physics processes beyond the Standard Model are placed as a
  function of the dilepton invariant mass within a fiducial region close to the experimental 
  selection criteria. The 95\% confidence level upper limits on the cross section of 
  anomalous $\ee$, $\emu$, or $\mm$ production range between 1.7~fb and 64~fb depending 
  on the dilepton mass and flavour combination. 
}

\newcommand{\hpp}{H^{\pm\pm}}

\begin{document}
\maketitle

\section{Introduction}
\label{sec:intro}

Events containing two prompt leptons with large transverse momentum ($\pt$) and equal electric charge are rarely produced in the Standard Model (SM) 
but occur with an enhanced rate in many models of new physics. For instance,
left-right symmetric models~\cite{LRSM1,LRSM2,LRSM3,LRSM4}, Higgs triplet 
models~\cite{Hpp_triplet,triplet1,triplet2}, the little Higgs 
model~\cite{hpp_little}, fourth-family quarks~\cite{theory4th},
supersymmetry~\cite{theorySUSY}, universal extra dimensions~\cite{theoryUED}, 
and the neutrino mass model of refs.~\cite{ZBM1,ZBM2,ZBM3} may produce final 
states with two like-sign leptons.  In the analysis described here, pairs of isolated, high-$\pt$
leptons are selected, and the invariant mass of the 
dilepton system ($e^\pm e^\pm$, $e^\pm\mu^\pm$, $\mu^\pm\mu^\pm$) is examined for the inclusive final state 
and separately for positively- and negatively-charged pairs.

The ATLAS Collaboration has previously reported inclusive searches for new physics in
the like-sign dilepton final state in a data sample corresponding to
an integrated luminosity of 34~$\ipb$~\cite{2010SS} and in like-sign muon pairs
with 1.6~fb$^{-1}$~\cite{2011SSmm}. No significant deviation from SM
expectations was observed, and fiducial production cross-section limits as well
as limits on several specific models of physics beyond SM were
derived. The CDF Collaboration has performed similar inclusive 
searches~\cite{cdf,cdf2} without observing any evidence for new physics.  Furthermore, the ATLAS and CMS Collaborations have performed several searches for like-sign leptons produced in association with jets or missing transverse momentum where no evidence of non-SM physics was observed~\cite{CMS_SS_b, CMS_SS_bprime, CMS_SS_METjet, CMS_SS_sstop, ATLAS_SS_METjet, ATLAS_SS_WRHN, ATLAS_SS_sstop, ATLAS_SS_bh}.

This article is organised as follows. A brief description of the ATLAS detector is given in section~\ref{sec:atlas}.
The data and simulation samples, the event selection, and the background determination are explained in sections~\ref{sec:datamc},
~\ref{sec:evsel}, and~\ref{sec:backgrounds}, respectively. The systematic uncertainties 
on the background estimate (including theoretical uncertainties on the production cross sections) and signal acceptances are summarised in section~\ref{sec:systematics}.
In section~\ref{sec:results} the number of observed lepton pairs in data is compared to the background estimate,
and in section~\ref{sec:limits} these results are used to derive upper limits on the fiducial
cross section for like-sign dilepton production in a kinematic region closely related to the 
experimental event selection.

\section{The ATLAS detector}
\label{sec:atlas}

The ATLAS detector~\cite{atlas} consists of an inner tracking system, 
calorimeters, and a muon spectrometer.
The inner detector (ID), directly surrounding the interaction point, is composed of a silicon pixel 
detector, a silicon microstrip detector, and a transition radiation tracker, all
immersed in a 2~T axial magnetic field. It covers the pseudorapidity\footnote{ATLAS uses a right-handed coordinate system with its origin 
at the nominal interaction point (IP) in the centre of the detector and the $z$-axis 
along the beam line. The $x$-axis points from the IP to the centre of the LHC ring, and 
the $y$-axis points upward. Cylindrical coordinates $(r,\phi)$ are used in the transverse $(x,y)$ 
plane, $\phi$ being the azimuthal angle around the beam line. The pseudorapidity is defined 
in terms of the polar angle $\theta$ as $\eta=-\ln\tan(\theta/2)$.} range 
$|\eta|<2.5$ and is enclosed by a calorimeter system consisting of electromagnetic and hadronic sections. 
The electromagnetic part is a lead/liquid-argon sampling calorimeter, divided into a 
barrel ($|\eta|<1.475$) and two end-cap sections ($1.375<|\eta|<3.2$). The barrel ($|\eta|<0.8$) 
and extended barrel ($0.8<|\eta|<1.7$) hadronic calorimeter sections consist of iron and scintillator 
tiles, while the end-cap ($1.5<|\eta|<3.2$) and forward ($3.1<|\eta|<4.9$) calorimeters are composed of copper or tungsten, and liquid-argon.

The calorimeter system is surrounded by a large muon 
spectrometer (MS) built with air-core toroids.
This spectrometer is equipped with precision tracking chambers (composed of monitored drift tubes and 
cathode strip chambers) to provide 
precise position measurements in the bending plane in the range $|\eta|< 2.7$.
In addition, resistive plate chambers and thin gap chambers with 
a fast response time are used primarily to trigger muons in the 
rapidity ranges $|\eta|\le 1.05$ and $1.05 <|\eta|< 2.4$, respectively.
The resistive plate chambers and thin gap chambers also
provide position measurements in the non-bending plane, which are used for the 
pattern recognition and the track reconstruction.

The ATLAS trigger system has a hardware-based Level-1 trigger followed by a software-based high-level trigger~\cite{triggerpaper}.
The Level-1 muon trigger searches for hit coincidences between different 
muon trigger detector layers inside geometrical windows that define the muon transverse 
momentum and provide a rough estimate of its position. It selects high-$\pt$ muons in the pseudorapidity range $|\eta|<2.4$.  
The Level-1 electron trigger selects local energy clusters of cells in the electromagnetic section
of the calorimeter. For low-energy electron clusters, low activity in the hadronic calorimeter 
nearby in the $\eta$-$\phi$ plane is required. 
The high-level trigger selection is
based on similar reconstruction algorithms as those used offline.

\section{Data sample and Monte Carlo simulation}
\label{sec:datamc}

This analysis is carried out using a data sample corresponding to an integrated 
luminosity of $4.7 \pm 0.2$ \ifb\ of $pp$ collisions, recorded in 2011 at a centre-of-mass energy of 7~TeV. 
In this dataset, the average number of interactions per beam crossing ranges from about six
in the first half of the year to about fifteen at the end of 2011.

The data were selected using single-muon and single-electron triggers 
with $\pt$ thresholds of 10~GeV and 16~GeV at Level-1, respectively. 
In the high-level trigger, a muon with $\pt>18$~GeV is required, while for electrons, the $\pt$ threshold is 20~GeV in the early 2011 data and 22~GeV in the later part of the year. 
To ensure no efficiency loss for electrons with very high-$\pt$, a trigger with a $\pt$ threshold of 45~GeV is 
also used which has no requirement on the hadronic calorimeter energy deposits near the electron in the $\eta$-$\phi$ plane 
at Level-1.

Monte Carlo (MC) simulation is used to estimate some of the background contributions 
and to determine the selection efficiency and acceptance for possible new physics signals. The dominant 
SM processes that contribute to prompt like-sign dilepton production
are $WZ$ and $ZZ$, with smaller contributions from like-sign $W$ pair production ($W^\pm W^\pm$) and production of a $W$ or $Z$ boson in association with a top quark pair ($t\bar{t}W$, and $t\bar{t}Z$). 
These are all estimated using MC simulation.  
For processes with a $Z$ boson, the contribution from $\gamma^*\to \ell^+\ell^-$ due to internal or external bremsstrahlung of final-state quarks or leptons is also 
simulated for $m(\ell^+\ell^-)>0.1$~GeV. 
\textsc{Sherpa}~\cite{sherpa} is used to generate $WZ$ and $ZZ$ events, while
$W^\pm W^\pm$, $t\bar{t}W$, and  $t\bar{t}Z$ production is modelled using {\sc MadGraph}~\cite{madgraph} 
for the matrix element and {\sc Pythia}~\cite{pythia} for the parton shower and fragmentation.  
The $W^\pm W^\pm$ sample includes the $W^\pm W^\mp W^\pm $ process.

The normalisation of the $WZ$ and $ZZ$ MC samples is based on cross sections determined at next-to-leading order (NLO) with {\sc MCFM}~\cite{mcfm}.
The cross sections times branching ratios for
$W^{\pm}Z \rightarrow \ell^{\pm}\nu\ell^{\pm}\ell^{\mp}$ and $ZZ \rightarrow
\ell^{\pm}\ell^{\mp}\ell^{\pm}\ell^{\mp}$ after requiring two charged leptons (electrons, muons, or taus)
with the same electric charge and with $\pt>20$~GeV and 
$|\eta|<2.5$, are $372$~fb and $91$~fb, respectively.

For $t\bar{t}W$ and $t\bar{t}Z$ production, the higher-order corrections are calculated in ref.~\cite{ttW} 
and refs.~\cite{ttZ1,ttZ2,ttZ3}, respectively. 
The full higher-order corrections for $W^{\pm}W^{\pm}$ production have not been calculated.  However, for parts of the process, the NLO QCD corrections have been shown to be small~\cite{nlowpwp,Melia:2010bm}.  Based on this, no correction is applied to the LO cross section.

Opposite-sign dilepton events due to $Z/\gamma^{*}$, $t\bar{t}$, and $W^\pm W^\mp$ production 
constitute a background if the charge of one of the leptons is misidentified. 
The $Z/\gamma^*$ process is generated with \textsc{Pythia}, and
the cross section is calculated at next-to-next-to-leading order (NNLO)
using PHOZPR~\cite{Hamberg:1990np}.
The ratio of this cross section to the leading-order cross section is used to determine a mass dependent QCD $K$-factor which is 
applied to the result of the leading-order simulation~\cite{Aad:2011xp}.  
Higher-order electroweak corrections (beyond the photon radiation included in the simulation) 
are calculated using \textsc{Horace}~\cite{horace,CarloniCalame:2007cd}, yielding an electroweak $K$-factor due to virtual heavy gauge-boson loops.
The production of $t\bar{t}$ is modelled using \textsc{MC@NLO}~\cite{mcatnlo}, with \textsc{Herwig}~\cite{herwig} used for parton showering and hadronisation. 
The normalisation is obtained from approximate NNLO QCD calculations using \textsc{Hathor}~\cite{Aliev:2010zk}. 
The production of $W^\pm W^\mp$ is generated using 
\textsc{Herwig}, and the cross section is normalised to the NLO value calculated 
with {\sc MCFM}. 
The production of $Z\gamma\to \ell\ell\gamma$ and $W\gamma\to \ell\nu\gamma$ constitutes a background if 
the photon converts to an $e^+e^-$ pair. {\sc Alpgen}~\cite{alpgen} is used to generate $W\gamma$ production,
and the cross section is normalised 
to be consistent with the ATLAS $W\gamma$ cross-section measurement for $\pt(\gamma)>15$~GeV~\cite{wgamma}.
The $Z\gamma$ process is implicitly accounted for in simulation of the $Z/\gamma^{*}$ process.

In addition, a variety of new physics signals are simulated in order to study the efficiency and 
acceptance of the selection criteria. Pair production of doubly-charged Higgs bosons 
($pp\to \hpp H^{\mp\mp}$) via a virtual $Z/\gamma^*$ exchange
is generated using {\sc Pythia} for $\hpp$ mass values between 50~GeV and 1000~GeV~\cite{theory_hpphmm}.
Production of a right-handed $W$ boson ($W_R$) 
decaying to a charged lepton and a right-handed neutrino ($N_R$)~\cite{theory_LRSM,theoryMNeu} 
and pair production of heavy down-type fourth generation quarks ($d_4$) decaying to $tW$
are generated using {\sc Pythia}. Like-sign top-quark pair production can occur in models 
with flavour-changing neutral currents, e.g.~via a $t$-channel exchange of a $Z'$ boson with $utZ'$ coupling. 
Since the left-handed coupling is highly constrained by $B^{0}_d-\bar{B}^{0}_d$ 
mixing~\cite{bdmix}, only right-handed top quarks ($t_R$) are considered. 
A sample is generated  with the {\sc Protos}~\cite{protos} generator, based on an effective four-fermion 
operator $uu \to tt$ corresponding to $Z'$ masses $\gg 1$~TeV~\cite{AguilarSaavedra:2011zy}.
The parton shower and hadronisation are performed with {\sc Pythia}. 

Parton distribution functions (PDFs) taken from CTEQ6L1~\cite{cteq6l1} are used for the {\sc MadGraph} and {\sc Alpgen} samples,
and the {\sc MRST2007 LO$^{**}$}~\cite{mrst} PDF set is used for the {\sc Pythia} and {\sc Herwig}  
samples. 
For the $t\bar{t}$ \textsc{MC@NLO} sample, CTEQ6.6~\cite{cteq66} PDFs are used, while for the generation of diboson samples with 
{\sc Sherpa} the CTEQ10~\cite{ct10} parameterisation is used. Uncertainties on the quoted cross sections are discussed in section~\ref{sec:systematics}.

The detector response to the generated events is simulated with the ATLAS simulation framework~\cite{atlassim} 
using \textsc{Geant4}~\cite{g4}, and the events are reconstructed with the same 
software used to process the data. Simulated minimum bias interactions generated with {\sc Pythia} are overlaid on the hard scatter events to 
closely emulate the multiple $pp$ interactions present in the current and in adjacent bunch crossings (pileup) present in the data. The simulated response is corrected for the small 
differences in efficiencies, momentum scales, and momentum resolutions observed between data and simulation.

\section{Event selection}
\label{sec:evsel}

Events are selected if they contain at least two leptons ($ee$, $e\mu$, $\mu\mu$) of the same electric charge with $\pt>20$~GeV.
If the higher-$\pt$ lepton is an electron, it is required to have $\pt>25$~GeV.
At least one muon with $\pt>20$~GeV or one electron with $\pt>25$~GeV is required to match
an object that passed the relevant high-level trigger. These $\pt$ thresholds are chosen such
that they are on the plateau of the trigger efficiency.
In addition to the lepton selection, events must have a primary vertex reconstructed
with at least three tracks
 with $\pt>0.4$~GeV.  If more than one interaction vertex is reconstructed in an event, the one with the highest
$\sum\pt^2$, summed over the tracks associated with the vertex, is chosen as the primary vertex.  For $Z$ bosons decaying to electrons or muons, the efficiency of the primary vertex requirements is close to 100\%.

Electrons are identified as compact showers in the electromagnetic sections of the calorimeters 
that are matched to a reconstructed track in the ID using the {\it tight}
criteria described in ref.~\cite{epaper}.  These criteria include requirements on the transverse shower shapes, the geometrical
match between the track and shower, the number and type of hits in the ID, and they reject electrons associated with reconstructed photon conversion vertices. Electrons must have $|\eta|<2.47$, excluding 
the transition region between the barrel and end-cap calorimeters ($1.37<|\eta|<1.52$). The electron tracks 
are refitted with the Gaussian Sum Filter algorithm to account for radiative energy losses (bremsstrahlung) 
in the detector material~\cite{GSF}.

Muon candidates are formed from tracks reconstructed in the ID combined with tracks reconstructed in the MS~\cite{staco}.
To reduce the charge mismeasurement rate the independent charge measurements from these two detectors are required to agree.  Muons must have $|\eta|<2.5$.

The ID tracks of electron and muon candidates are required to be consistent with 
originating from the primary interaction vertex. The requirements are $|d_0|<1.0$~mm,
$|z_0\sin\theta|<1.0$~mm, and $|d_0/\sigma(d_0)|<3$, where $d_0$ ($z_0$) is the transverse (longitudinal) distance between the primary vertex and the point of closest approach of the track and
$\sigma(d_0)$ is the 
estimated uncertainty on the $d_0$ measurement, typically  $15$~$\mu$m. 
To ensure precise impact-parameter measurements, a hit in the innermost pixel detector 
layer is required if the track traverses an active detector component.
This requirement also reduces background due to electrons from photon conversions. 

Both electrons and muons are required to be isolated from other activity in the event. The isolation 
energy, $E_{\rm T}^{{\rm cone} \Delta R_{\rm iso}}$, is the sum of transverse energies\footnote{Transverse energy is the projection of an energy deposition in the calorimeter in the transverse plane, where the direction is given by the line joining the energy deposition and the nominal interaction point.} in calorimeter cells 
(including electromagnetic and hadronic sections) in a cone $\Delta R=\sqrt{(\Delta\eta)^2+(\Delta\phi)^2}<\Delta R_{\rm iso}$
around the lepton direction.  This quantity is corrected for the energy of the electron object as well as 
energy deposits from pileup interactions. 
The track isolation, $p_{\rm T}^{{\rm cone} \Delta R_{\rm iso}}$, is analogously defined as the scalar sum of the transverse
momentum of tracks with $\pt>1$~GeV in a cone of $\Delta R<\Delta R_{\rm iso}$ around the lepton direction, excluding the lepton track. 
For muons, $p_{\rm T}^{\rm cone0.4}/\pt(\mu)<0.06$ and $p_{\rm T}^{\rm cone0.4}<4\mbox{ GeV}+0.02\times \pt(\mu)$ are required.
For electrons, the requirements are $p_{\rm T}^{\rm cone0.3}/\pt(e)<0.1$ and  
$E_{\rm T}^{\rm cone0.2}<3\mbox{ GeV}+(\pt(e)-20\mbox{ GeV}) \times 0.037$.  These requirements are chosen to maintain good efficiency for prompt leptons over a large $\pt$ range while rejecting a large fraction of non-prompt leptons, particularly at low $\pt$.

In addition to the isolation criteria, selected leptons are required to be well separated from jets to suppress leptons from hadronic
decays. For this purpose, jet candidates are reconstructed from 
topological clusters in the calorimeter~\cite{topocluster} using the anti-$k_t$ algorithm~\cite{antikt} with a radius parameter of 0.4.
Their energies are corrected for calorimeter non-compensation, energy loss in upstream material, and other 
instrumental effects. Furthermore, quality criteria are applied to remove reconstructed jets not arising
from hard-scattering interactions~\cite{jetcal}. Jets are required to have $\pt>25$~GeV, $|\eta|<2.8$, and 75\% of the 
momentum of the tracks within the jet must originate from the primary vertex. 
Any jet within a distance $\Delta R = 0.2$ of a candidate electron is removed from the list of jets to avoid counting the electron 
as a jet. Electrons and muons are then required to be separated by $\Delta R>0.4$ from any jet 
with $\pt>25\mbox{ GeV} + 0.05\times \pt(\ell)$. 

Lepton pairs are selected if they contain two leptons with the same electric charge passing the above selection requirements 
with invariant mass $m(\ell^{\pm}\ell^{\pm}) > 15$~GeV. Furthermore, for $e^\pm e^\pm$ pairs, the mass  
range 70--110~GeV is vetoed due to large backgrounds from opposite-sign electron pairs produced by $Z$ decays, where the charge of one lepton 
is misidentified.  Any combination of two leptons is considered, allowing more than one lepton pair per event to be included.

\section{Background determination}
\label{sec:backgrounds}

Backgrounds to this search arise from three principal sources: SM production of prompt, like-sign lepton pairs 
(prompt background); production of opposite-sign lepton pairs where the charge of one lepton is 
misidentified (charge-flip background) or where a photon produced in association with the opposite-sign leptons converts into an $e^+e^-$ pair; 
and hadrons or leptons from hadronic decays (non-prompt background).

\subsection{Prompt lepton backgrounds}
\label{sec:prompt} 
SM processes resulting in two prompt leptons of the same electric charge are $WZ$, $ZZ$, $W^\pm W^\pm$, $t\bar{t}W$, and 
$t\bar{t}Z$ production.  The background due to these processes is determined from MC simulation using the 
samples and cross sections described in section~\ref{sec:datamc}.  Other SM processes are not expected to contribute significantly to the background and are neglected.  

\subsection{Charge misidentification and photon conversions}
\label{sec:chargeflip} 
Charge misidentification can occur for high-momentum tracks when the tracking detector is unable to 
determine the curvature of the track. This misidentification can occur for both electrons and muons. In the momentum range of a few hundred GeV 
relevant for this analysis, the background arising from this source is determined to be negligible based on MC simulation. 
The rate of misidentification is studied in data with 
muons from $Z\to\mu\mu$ decays where the muons are selected based only on the MS (ID) measurements to study the 
ID (MS) mismeasurement. The combined probability to mismeasure the charge in both the ID and the MS is found to 
be consistent with zero and an upper limit ranging from $5 \times 10^{-10}$ at low $\pt$ to $7 \times 10^{-2}$ 
for $\pt \sim 400$~GeV is placed. This upper limit is used to determine the systematic uncertainty for muons. 

Another source of charge misidentification occurs for electrons when a
high-momentum photon is radiated and converts into an $e^+e^-$ pair.
Electron candidates are rejected if they are associated with conversion vertices~\cite{epaper}.
However, for asymmetric conversions, 
sometimes only one of the tracks is reconstructed and its  
charge may be different from the charge of the original electron that radiated the photon.
The probability for electron charge misidentification to occur, either by this mechanism or by track mismeasurement, is measured as the fraction
of like-sign $e^\pm e^\pm$ pairs with 80~GeV$< m(ee) <100$~GeV as a function of the electron $\eta$
following the procedure explained in ref.~\cite{atlasSStop}. 
The simulation is found to overestimate this probability by about 15\%. An $\eta$-dependent scaling factor is applied to the simulated rate 
to account for this overestimate, and the simulation is then used 
to predict the backgrounds from $Z/\gamma^{*}$, $t\bar{t}$, and $W^\pm W^\mp$ production. In the central (forward)
region the probability increases from about 0.01\% (0.6\%) at $\pt\approx 50$~GeV to 0.3\% (4\%) at $\pt\approx 300$~GeV. 
An uncertainty on this misidentification probability of $\pm$(10--20)$\%$ (depending on $\eta$) is derived by comparing 
different methods to determine this factor as described in ref.~\cite{atlasSStop} and by considering the statistical uncertainty on each method.  

Production of $W\gamma\to\ell\nu\gamma$ and $Z\gamma\to \ell^+\ell^-\gamma$ can lead to like-sign lepton pairs
if the photon converts. This background is determined using the MC samples described in section~\ref{sec:datamc}.
Since this background is closely related to the charge misidentification for electrons described above, the 
MC-based estimate is also scaled down by 15\% and the same systematic uncertainty of $\pm$(10--20)$\%$ is applied.

\subsection{Non-prompt lepton backgrounds}
\label{sec:nonprompt}

The non-prompt background includes lepton pairs where one or both of the leptons result from hadronic decay or misidentification.  Processes contributing to this background include $W+$jet, $Z+$jet, multi-jet (including $b\bar{b}$ and $c\bar{c}$), and $t\bar{t}$ production.

The non-prompt background is determined directly from data. For both electrons and muons, the determination relies on measuring a
factor $f$ that is the ratio of the number of selected leptons satisfying the analysis selection criteria ($N_S$) to the number 
of leptons failing these selection criteria but passing a less stringent set of requirements, referred to as anti-selected leptons ($N_A$): 

$$f=\frac{N_S}{N_A}.$$
The factor $f$ is determined as a function of $\pt$ and $\eta$ in a data sample dominated by non-prompt leptons. 
Any contamination by prompt leptons in the numerator and denominator is subtracted
using MC simulation.  The details of the anti-selected definitions and the regions used for measuring $f$ depend on the lepton flavour and are described in more detail below.

The primary sources of non-prompt muons are semi-leptonic
decays of $b$- and $c$-hadrons. 
For the anti-selection, the isolation criterion described in section~\ref{sec:evsel} is inverted, but a looser 
requirement of $p_{\rm T}^{\rm cone0.4}/\pt(\mu)<1.0$ is placed.
The measurement of $f$ is done in a sample dominated by non-prompt muons with $|d_0|/\sigma(d_0)>5$ and $|d_0|<10$~mm.
These muons are selected from dimuon events with $\pt(\mu)>10$~GeV, $m(\mu\mu)>15$~GeV, and where $Z$-boson candidates have been vetoed.
Simulated muons from $b$- and $c$-hadron decays are used to determine the expected difference between $f$ for muons with $|d_0|/\sigma(d_0)>5$ (used in the data-driven determination of $f$) and muons with $|d_0|/\sigma(d_0)<3$ (used in the event selection).  In simulation, muons with $|d_0|/\sigma(d_0)<3$ are more isolated, so a correction factor of $1.34 \pm 0.34$ is applied to $f$.
The resulting value of $f$ is about 
0.10 at $\pt=20$~GeV and increases to about 0.25 at $\pt=100$~GeV. 
One source of uncertainty in this procedure is that the fraction of non-prompt muons from pion and kaon decays may be larger for muons passing the analysis-level impact parameter criteria than for those used in the measurement of $f$.  
The uncertainty due to the contribution of light-flavour decays is assessed by exploiting the
difference in the track momentum measurement in the ID and the MS, which is expected to be large for kaons or pions
decaying in the ID or calorimeters.
The difference between $f$ for light-flavour decays and heavy-flavour decays is combined with the fraction of non-prompt muons from light-flavour decays to give an uncertainty of $\pm15\%$ on $f$.
The total systematic uncertainty on $f$ is the quadratic sum of the statistical error, the uncertainty on the 
prompt background subtraction, the correction for the isolation-dependence on the impact parameter significance criteria,
and the uncertainty in the contribution of pion and kaon decays. It is about $\pm37\%$ at low $\pt$ and $\pm100\%$ for $\pt>100$~GeV.

For electrons, the main non-prompt backgrounds arise from heavy-flavour decays, charged pions that shower early in the calorimeter,
and neutral pions decaying to two photons where one of the photons converts to an $e^+e^-$ pair. 
The anti-selected electrons must have either $|d_0|/\sigma(d_0)>3$ or fail the {\it medium} electron
identification criteria while passing the {\it loose} electron criteria, as defined in ref.~\cite{epaper}.
The inversion of the impact parameter criteria enhances the heavy-flavour background, while the inversion of the 
{\it medium} identification criteria enhances the pion backgrounds.
A sample dominated by non-prompt electrons is selected by triggering on a high-$\pt$ electron candidate and 
requiring that the event contains a jet with $\pt >20$~GeV. Contamination of this sample by prompt electrons is reduced by requiring
that there be only one electron candidate and requiring the transverse mass\footnote{The transverse mass is defined as
$\sqrt{2E_{\rm T}^{\rm miss}p_{\rm T}(e)(1-\cos\Delta\phi(e,E_{\rm T}^{\rm miss}))}$, where $E_{\rm T}^{\rm miss}$ 
is the missing transverse momentum and $\Delta\phi(e,E_{\rm T}^{\rm miss})$ is the difference between the azimuthal angles of the
electron and $E_{\rm T}^{\rm miss}$.} formed by the electron and the missing transverse momentum ($\met$) 
to be below 40~GeV. 
The value of $f$ is found to be about 0.18 at $\et=20$~GeV and decreases to about 
0.10 at $\et=100$~GeV. A systematic uncertainty is determined accounting for the statistical uncertainty, the prompt
background subtraction, 
the $\pt$ requirement of the jet in the event, and any trigger bias. 
The uncertainty is about $\pm10\%$ at low $\et$ and increases to $\pm100\%$ for $\et>300$~GeV.  Alternate anti-selection definitions are 
used to assess the overlap between the non-prompt and charge misidentification backgrounds, as well as the contributions of 
light- and heavy-flavour jets to the non-prompt background.  For the former, a correction is made to the non-prompt background, 
amounting to $\pm$(11--25)$\%$, and the full correction is taken as a systematic uncertainty. 
For the latter, an uncertainty of $\pm$(4--84)$\%$ is found, depending on the invariant mass threshold considered.

A background prediction is derived from $f$ using
dilepton pairs where one or both leptons are anti-selected but pass all other event selection criteria.
The non-prompt background prediction is then given by
\begin{equation}
N_{\text{non-prompt}}= \sum_{i=1}^{N_{A+S}} f(\ptone,\eta_1) + \sum_{i=1}^{N_{S+A}} f(\pttwo,\eta_2) - \sum_{i=1}^{N_{A+A}} f(\ptone,\eta_1) f(\pttwo,\eta_2),
\end{equation} 
where the sums are over the number of pairs, $N_{A+S}$ ($N_{S+A}$), where the subleading (leading) lepton satisfies the selection criteria and the other lepton satisfies the anti-selection, or over the number of pairs where both leptons are anti-selected ($N_{A+A}$).  The prompt contributions to $N_{A+S}$, $N_{S+A}$, and $N_{A+A}$ are subtracted based on MC predictions.
The variables $\ptone$ and $\eta_1$ ($\pttwo$ and $\eta_2$) refer to the kinematic properties of the 
leading (subleading) lepton.

The background estimates are verified in several control regions that 
are designed to probe specific sources of background. The understanding of prompt leptons is tested in a sample requiring
two leptons of opposite charge. This sample is dominated by the $Z/\gamma^{*}$ process in all three final states, and the data agree with the background prediction to better than 5\%. 
The kinematics of the electron charge misidentification background are verified in pairs of like-sign electrons with 80~GeV~$<m(e^{\pm}e^{\pm})<100$~GeV.
The non-prompt background 
is tested by inverting the isolation criteria but requiring a looser isolation cut value.
The heavy-flavour background is tested by inverting the $|d_0|/\sigma(d_0)$ criterion. Furthermore, for electrons, the light-flavour 
background is tested by inverting some of the shower shape criteria. The 
data agree with the background prediction in the various control regions within the systematic uncertainties.

\section{Systematic uncertainties}
\label{sec:systematics}

Several systematic effects can change the signal acceptance and the background estimate. 
Uncertainties on the event selection efficiencies and on the luminosity affect the predicted
yield of signal events as well as those backgrounds that are estimated from MC simulation.  Uncertainties on the trigger efficiencies, lepton identification efficiencies,
 and lepton momentum scales are determined using $W$, $Z$, and $J/\psi$ decays~\cite{electronPerf, muonMomRes, muonEff}.
The uncertainties on the lepton identification efficiencies 
result in an uncertainty on the number of like-sign pairs of $\pm$(3--4)$\%$. Effects resulting from uncertainties in the muon momentum scale and resolution are negligible.
For electrons the energy scale is known to $\pm 1\%$, and 
its impact on the signal acceptance is below 0.1\%, while the background estimate is affected by up to $\pm 3.5\%$. 
The uncertainties due to the trigger efficiencies are $<1\%$ in all channels.
The uncertainty on the integrated luminosity is $\pm3.9\%$~\cite{lumipaper,atlaslumi2011}.

The uncertainties on the production cross sections of the SM processes
affect the predicted yield of the prompt lepton background. For $WZ$ and $ZZ$ production, an uncertainty of $\pm 10\%$ is estimated
due to higher-order QCD corrections by varying the renormalisation and factorisation scales by a factor of two.
The uncertainty due to the parton distribution functions (PDFs) is evaluated using the eigenvectors
provided by the CTEQ10 set~\cite{ct10} of PDFs,
following the prescription given in ref.~\cite{pdferr_new}. The difference between the central cross-section values obtained using this PDF set and that obtained with the MRST2008NLO PDFs~\cite{mrst} is also added in quadrature.  This procedure gives a conservative estimate of the PDF uncertainty on the cross section.
The resulting uncertainty on the cross section is $\pm 7\%$.
For $t\bar{t}W$, $t\bar{t}Z$ and $W^\pm W^\pm$ production a normalisation uncertainty of $\pm 50\%$ is assigned~\cite{ttZ1, nlowpwp}.

The production of $Z/\gamma^{*}$, $t\bar{t}$, and $W^\pm W^\mp$ also constitutes a background
in the $ee$ and $e\mu$ channels due to charge misidentification. 
The PDF uncertainties for $Z/\gamma^{*}$ and $t\bar{t}$ are computed using the MSTW2008 NNLO PDF sets~\cite{mstw2008} and the renormalisation and factorisation scales are varied by factors of two to derive the uncertainties due to higher-order QCD corrections.  For $Z/\gamma^{*}$, an additional systematic uncertainty is attributed to electroweak corrections~\cite{Aad:2011xp}.  The total uncertainties on the $Z/\gamma^{*}$ and $t\bar{t}$ cross sections are $\pm7\%$ and +10/--11\%, respectively.
The $W^\pm W^\mp$ cross-section uncertainty is determined using the same scale and PDF variations as for $WZ$ and $ZZ$, yielding a total uncertainty of $\pm12$\%.

The systematic uncertainty on the background from non-prompt leptons is discussed in section~\ref{sec:nonprompt}. 
It includes a component due to the systematic uncertainties associated with the measurement of the ratio $f$ and 
a component due to the statistical uncertainty on the number of anti-selected leptons.
For $m(\ell^{\pm}\ell^{\pm})>15$~GeV it is $\pm33\%$ for the $\mu\mu$, $\pm31\%$ for the $e\mu$, 
and $\pm28\%$ for the $ee$ final states, and it increases to nearly $\pm100\%$ at higher masses for all final states. 

The probability of assigning the wrong charge to the lepton and its uncertainty is discussed in section~\ref{sec:chargeflip}.
For the $\mu\mu$ final state, charge misidentification results in an uncertainty of $^{+4.9}_{-0.0}$ events for $m(\ell^{\pm}\ell^{\pm})>15$~GeV and 
$^{+1.7}_{-0.0}$ events for $m(\ell^{\pm}\ell^{\pm})>400$~GeV. For electrons, the uncertainty on the charge-flip background due to uncertainty in the charge mismeasurement probability ranges  from $\pm 15$\% to $\pm 23$\% for the same mass thresholds.
The uncertainty on the $W\gamma$ and $Z\gamma$ backgrounds from this source is $\pm$(15--18)$\%$ depending on $m(\ell^{\pm}\ell^{\pm})$.

Statistical uncertainties due to the limited size of the background MC samples are also considered.
Systematic uncertainties on different physics processes due to a given source are assumed to be 100\% correlated.

\section{Comparison of data to the background expectation}
\label{sec:results}

The predicted numbers of background pairs are compared to the observed numbers of like-sign lepton pairs 
in table~\ref{tab:2iso_SS_both}. For the $\ee$ ($\emu$) final state, about 30\% (50\%) of the background is from prompt leptons in all mass bins.  The rest arises from leptons from hadronic decays, charge flips, or photon conversions.
For the $\mm$ final state, the prompt background constitutes about 83\% of the total background for $m(\mm)>15$~GeV, 
rising to nearly 100\% of the total background for $m(\mm)>400$~GeV.
The overall uncertainty on the background is about $\pm15\%$ at low mass and $\pm30\%$ at high mass.

\begin{table*}[tbp]
\caption{Expected and observed numbers of pairs of isolated like-sign leptons for various 
cuts on the dilepton invariant mass, $m(\ell^{\pm}\ell^{\pm})$. The uncertainties shown are the quadratic 
sum of the statistical and systematic uncertainties.
The prompt background contribution includes the $WZ$, $ZZ$, $W^{\pm}W^{\pm}$, $t\bar{t}W$, and $t\bar{t}Z$ processes.  When zero events are predicted, the uncertainty corresponds to the 68\% confidence level upper limit on the prediction.}
\begin{center}
\begin{tabular}{l|c|c|c|c|c}
\hline
Sample & \multicolumn{4}{|c}{Number of electron pairs with  $m(e^{\pm}e^{\pm})$} \\
 & $>15$~GeV & $>100$~GeV & $>200$~GeV & $>300$~GeV & $>400$~GeV \\
\hline \hline
Prompt & $101 \pm 13$ & $56.3 \pm 7.2$ & $14.8 \pm 2.0$ & $4.3 \pm 0.7$ & $1.4 \pm 0.3$ \\[+0.05in]
Non-prompt      & $75 \pm 21$ & $28.8 \pm 8.6$ & $5.8 \pm 2.5$ & $0.5^{+0.8}_{-0.5}$ & $0.0^{+0.2}_{-0.0}$ \\[+0.05in]
Charge flips and & \multirow{2}{*}{$170 \pm 33$} & \multirow{2}{*}{$91 \pm 16$} & \multirow{2}{*}{$22.1 \pm 4.4$} & \multirow{2}{*}{$8.0 \pm 1.7$} & \multirow{2}{*}{$3.4 \pm 0.8$} \\
conversions & & & & & \\
\hline
Sum of backgrounds & $346 \pm 44$ & $176 \pm 21$ & $42.8 \pm 5.7$ & $12.8 \pm 2.1$ & $4.8 \pm 0.9$ \\[+0.05in]
\hline
Data       &        329 &        171 &         38 &         10 &          3 \\[+0.05in]
\hline
& \multicolumn{4}{|c}{Number of muon pairs with  $m(\mu^{\pm}\mu^{\pm})$} \\
 & $>15$~GeV & $>100$~GeV & $>200$~GeV & $>300$~GeV & $>400$~GeV \\
\hline \hline
Prompt & $205 \pm 26$ & $90 \pm 11$ & $21.8 \pm 2.8$ & $5.8 \pm 0.9$ & $2.2 \pm 0.4$ \\[+0.05in]
Non-prompt & $42 \pm 14$ & $12.1 \pm 4.6$ & $1.0 \pm 0.6$ & $0.0^{+0.3}_{-0.0}$ & $0.0^{+0.3}_{-0.0}$ \\[+0.05in]
Charge flips & $0.0^{+4.9}_{-0.0}$ & $0.0^{+2.5}_{-0.0}$ & $0.0^{+1.8}_{-0.0}$ & $0.0^{+1.7}_{-0.0}$ & $0.0^{+1.7}_{-0.0}$ \\[+0.05in]
\hline
Sum of backgrounds & $247^{+30}_{-29}$ & $102 \pm 12$ & $22.8^{+3.4}_{-2.9}$ & $5.8^{+1.9}_{-0.9}$ & $2.2^{+1.7}_{-0.4}$ \\[+0.05in]
\hline 
Data       &    264 &   110 &     29 &         6  & 2\\[+0.05in]
\hline
& \multicolumn{4}{|c}{Number of lepton pairs with  $m(e^{\pm}\mu^{\pm})$} \\
 & $>15$~GeV & $>100$~GeV & $>200$~GeV & $>300$~GeV & $>400$~GeV \\
\hline \hline
Prompt & 346 $\pm$ 43 & 157 $\pm$ 20 & 36.6 $\pm$ 4.7 & 10.8 $\pm$ 1.5 & 3.9 $\pm$ 0.6\\[+0.05in]
Non-prompt & 151 $\pm$ 47 & 45 $\pm$ 13 & 9.2 $\pm$ 4.1 & 2.6 $\pm$ 1.1 & 1.0 $\pm$ 0.6\\[+0.05in]
Charge flips and & \multirow{2}{*}{142 $\pm$ 28} & \multirow{2}{*}{33 $\pm$ 7} & \multirow{2}{*}{10.5 $\pm$ 2.8} & \multirow{2}{*}{2.9 $\pm$ 1.2} & \multirow{2}{*}{2.2 $\pm$ 1.1}\\
conversions & & & & & \\
\hline
Sum of backgrounds
 & 639 $\pm$ 71 & 235 $\pm$ 25 & 56.4 $\pm$ 7.0 & 16.3 $\pm$ 2.3 & 7.0 $\pm$ 1.4\\[+0.05in]
\hline
Data & 658 & 259 & 61 & 17 & 7\\[+0.05in]
\hline
\hline
\end{tabular}
\end{center}
\label{tab:2iso_SS_both}
\end{table*}

Table~\ref{tab:2iso_SS_sep} shows the data compared to the background expectation separately for 
$\ell^+\ell^+$ and $\ell^-\ell^-$ events for the $ee$, $\mu\mu$, and $e\mu$ final states. 
The background is higher for the $\ell^+\ell^+$ final state 
due to the larger cross section in $pp$ collisions for $W^+$ than $W^-$ bosons produced in association with $\gamma$, $Z$, or hadrons.

\begin{table*}[htbp]
\caption{Expected and observed numbers of positively- and negatively-charged lepton pairs for different 
lower limits on the dilepton invariant mass, $m(\ell^{\pm}\ell^{\pm})$. The uncertainties shown are the quadratic 
sum of the statistical and systematic uncertainties.}
\begin{center}
\begin{tabular}{l|c|c|c|c|c}
\hline
Sample & \multicolumn{4}{|c}{Number of lepton pairs with $m(\ell^\pm\ell^\pm)$} \\
 & $>15$~GeV & $>100$~GeV & $>200$~GeV & $>300$~GeV & $>400$~GeV \\
\hline 
& \multicolumn{4}{|c}{$e^+e^+$ pairs} \\\hline
Sum of backgrounds & $208 \pm 28$ & $112 \pm 14$ & $28.6 \pm 4.0$ & $8.5 \pm 1.4$ & $3.3 \pm 0.7$ \\[+0.05in]
Data       &        183 &         93 &         26 &          6 &          1 \\[+0.05in]
\hline\hline
& \multicolumn{4}{|c}{$e^-e^-$ pairs} \\\hline
Sum of backgrounds & $138 \pm 21$ & $63.3 \pm 8.5$ & $14.2 \pm 2.3$ & $4.4 \pm 0.8$ & $1.54^{+0.4}_{-0.3}$ \\[+0.05in]
Data       &        146 &         78 &         12 &          4 &          2 \\[+0.05in]
\hline\hline
& \multicolumn{4}{|c}{$\mu^+\mu^+$ pairs} \\\hline
Sum of backgrounds & $147 \pm 17$ & $63.7^{+7.7}_{-7.6}$ & $14.5^{+2.1}_{-1.9}$ & $4.1^{+1.1}_{-0.6}$ & $1.6^{+0.9}_{-0.3}$ \\[+0.05in]
Data       &        144 &        60 &         16 &         4 & 2 \\[+0.05in]
\hline\hline
& \multicolumn{4}{|c}{$\mu^-\mu^-$ pairs} \\\hline
Sum of backgrounds & $100 \pm 12$ & $38.4^{+5.0}_{-4.8}$ & $8.3^{+1.5}_{-1.2}$ & $1.7^{+0.9}_{-0.3}$ & $0.6^{+0.9}_{-0.1}$ \\[+0.05in]
Data       &        120 &        50 &         13 &         2 & 0 \\[+0.05in]
\hline\hline
& \multicolumn{4}{|c}{$e^+\mu^+$ pairs} \\\hline
Sum of backgrounds & 381 $\pm$ 42 & 142 $\pm$ 15 & 33.8 $\pm$ 5.3 & 9.8 $\pm$ 1.5 & 4.2 $\pm$ 0.9\\[+0.05in]
Data & 375 & 149 & 39 & 9 & 4\\[+0.05in]
\hline\hline
& \multicolumn{4}{|c}{$e^-\mu^-$ pairs} \\\hline
Sum of backgrounds & 259 $\pm$ 31 & 93 $\pm$ 10 & 22.6 $\pm$ 3.0 & 6.5 $\pm$ 1.3 & 2.9 $\pm$ 1.0\\[+0.05in]
Data & 283 & 110 & 22 & 8 & 3\\[+0.05in]
\hline\hline
\end{tabular}
\end{center}
\label{tab:2iso_SS_sep}
\end{table*}

The level of agreement between the data and the background expectation is evaluated using $1-$CL$_{b}$~\cite{cls}, defined 
as the one-sided probability of the background-only hypothesis to fluctuate up to at least the number of observed events.
Statistical and systematic uncertainties and their correlations 
are fully considered for this calculation. 
The largest upward deviation, observed for $m(\mu^-\mu^-)>100$~GeV, occurs about 8\% of the time in background-only pseudo-experiments for this mass bin.

Figure~\ref{fig:dilep_mass_SS} shows the dilepton invariant mass spectra for the $\ee$, $\mm$, and $\emu$ final states,
and figure~\ref{fig:dilep_pt_SS} shows the $\pt$ of the leading leptons. 
Good agreement with the background expectation is observed for all three channels. 

\begin{figure}[htbp]
\begin{center}
\subfigure[] {
  \includegraphics[width=0.46\columnwidth]{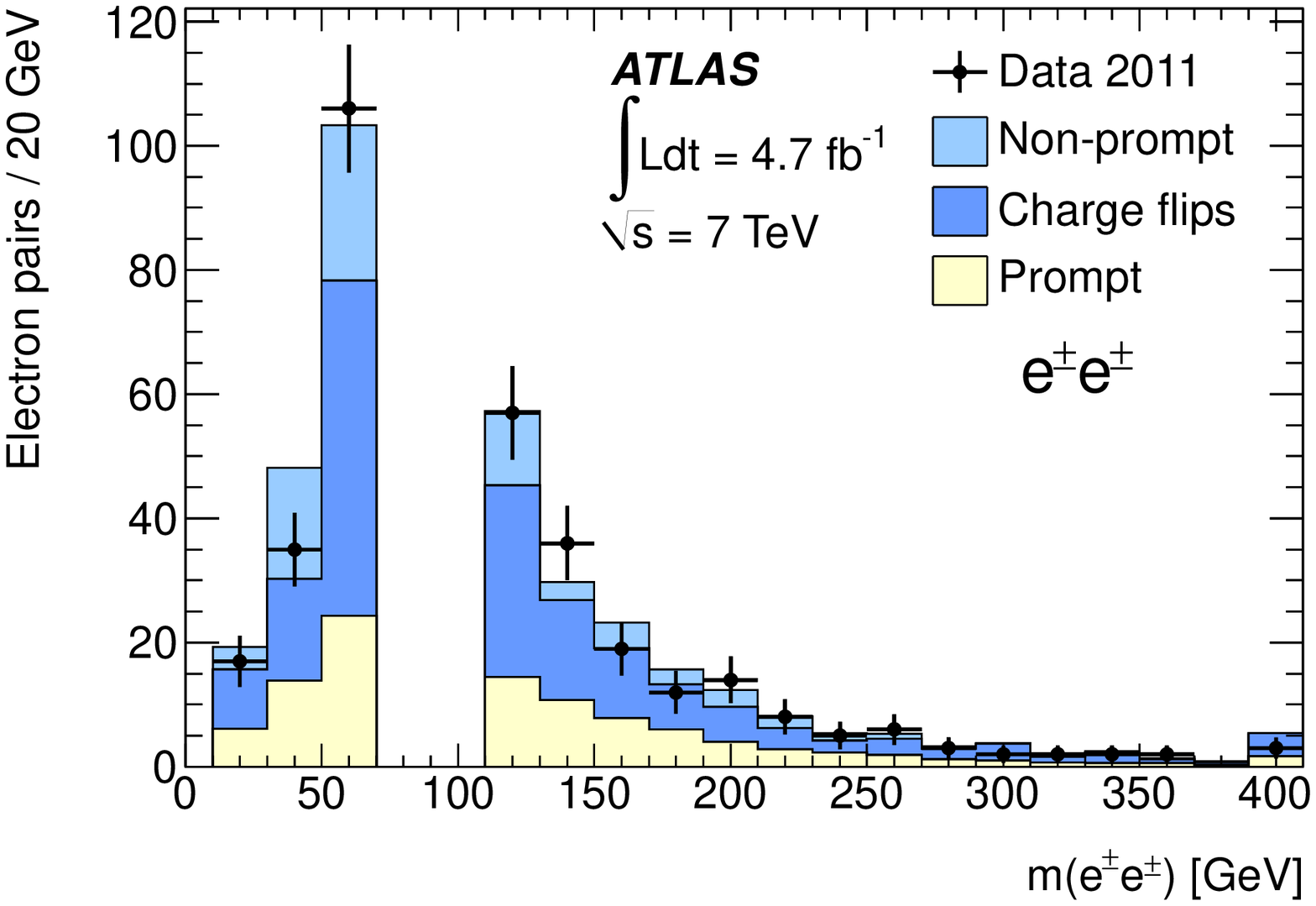}
}
\subfigure[] {
  \includegraphics[width=0.46\columnwidth]{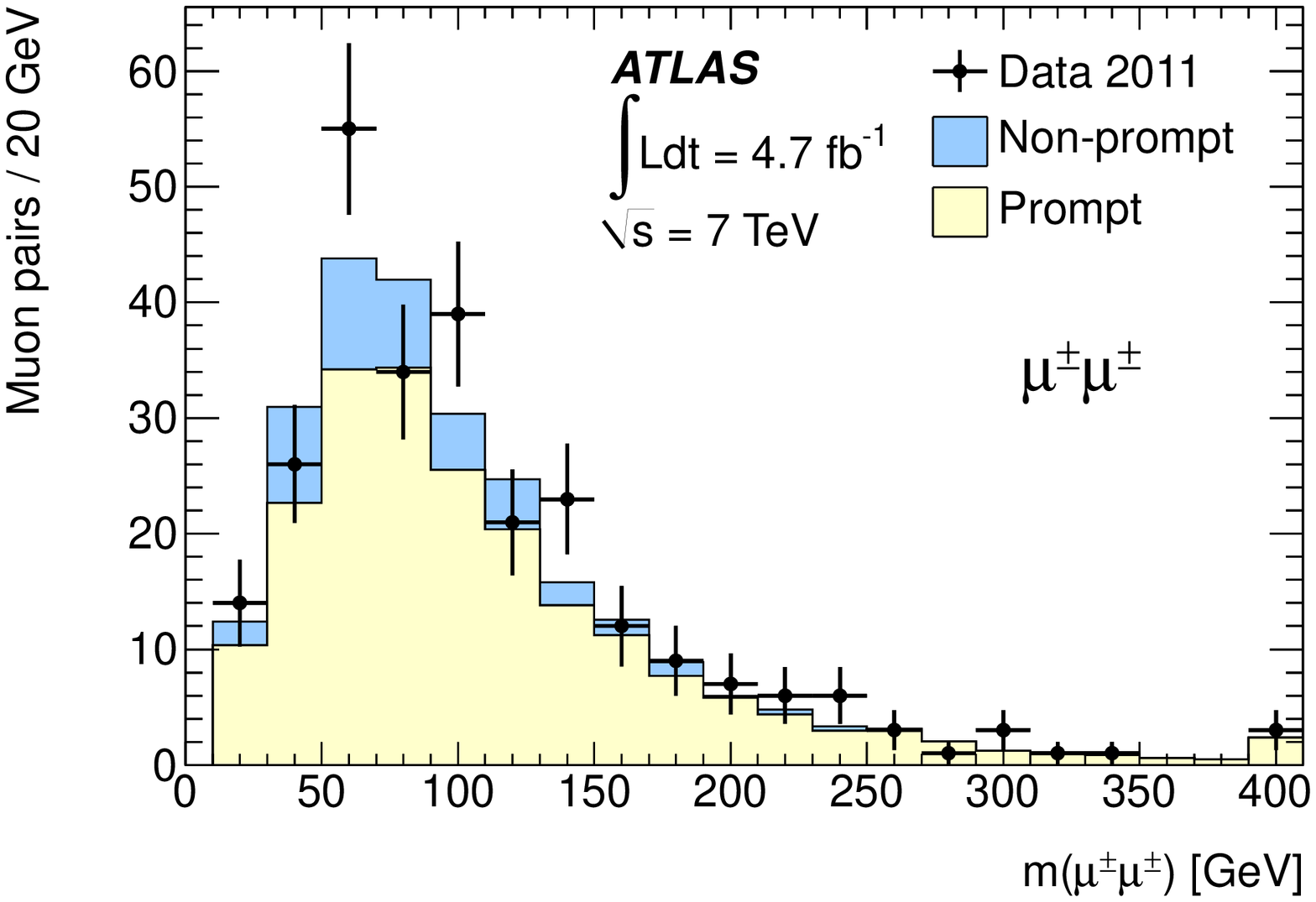}
}
\subfigure[] {
  \includegraphics[width=0.46\columnwidth]{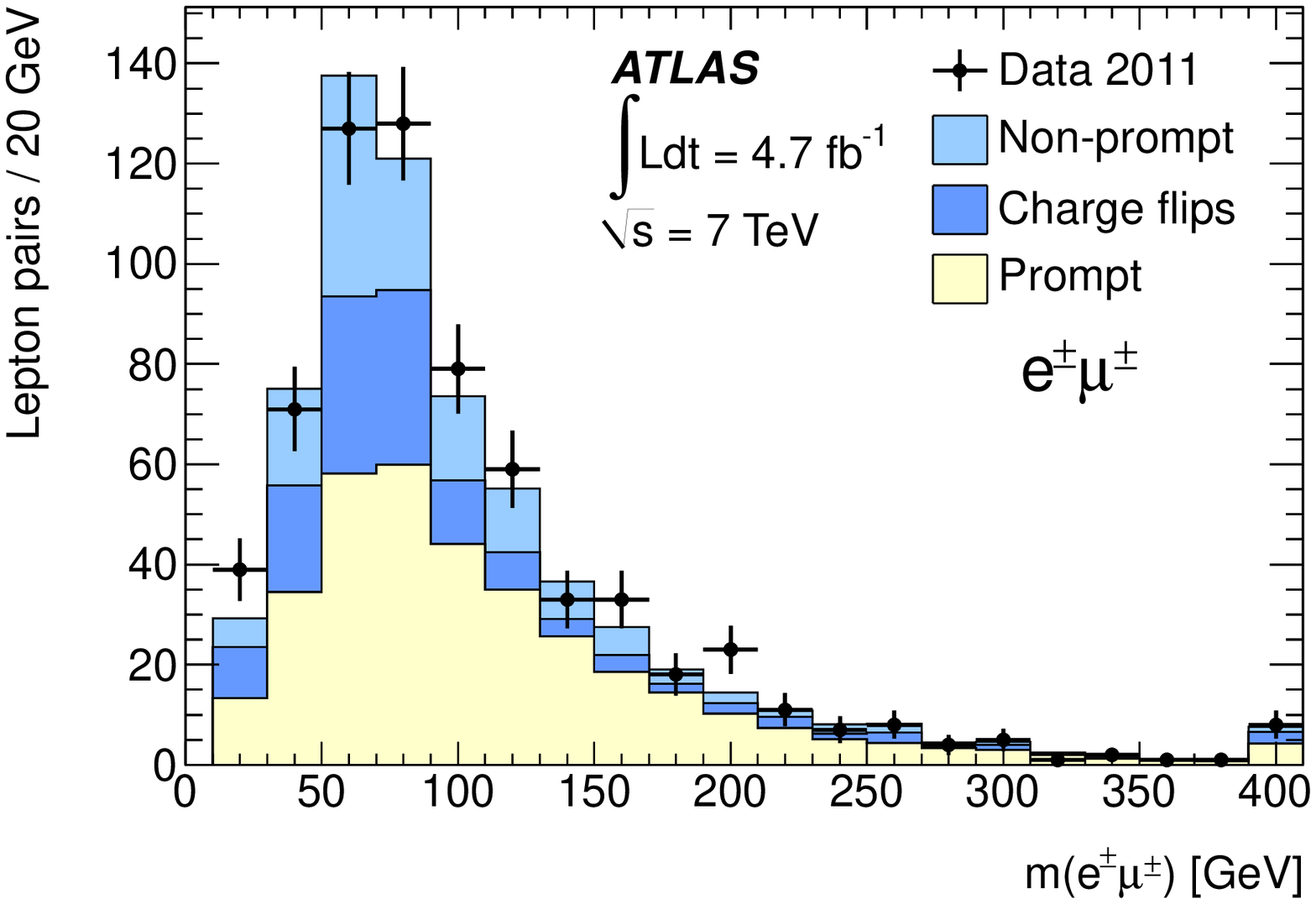}
}
\caption{\label{fig:dilep_mass_SS} Invariant mass distributions for (a) $\ee$, (b) $\mm$, and (c) $\emu$
pairs passing the full event selection. The data are shown as closed circles.  The stacked histograms represent the backgrounds composed of pairs of prompt leptons from SM processes, pairs with at least one non-prompt lepton, and for the electron channels, backgrounds arising from charge misidentification and photon conversions. Pairs in the $ee$ channel with invariant masses between 70~GeV and 110~GeV are excluded because of the large background from charge misidentification in $Z\to e^{\pm}e^{\mp}$ decays.  The last bin is an overflow bin.} 
\end{center}
\end{figure}

\begin{figure}[htbp]
\begin{center}
\subfigure[] {
  \includegraphics[width=0.46\columnwidth]{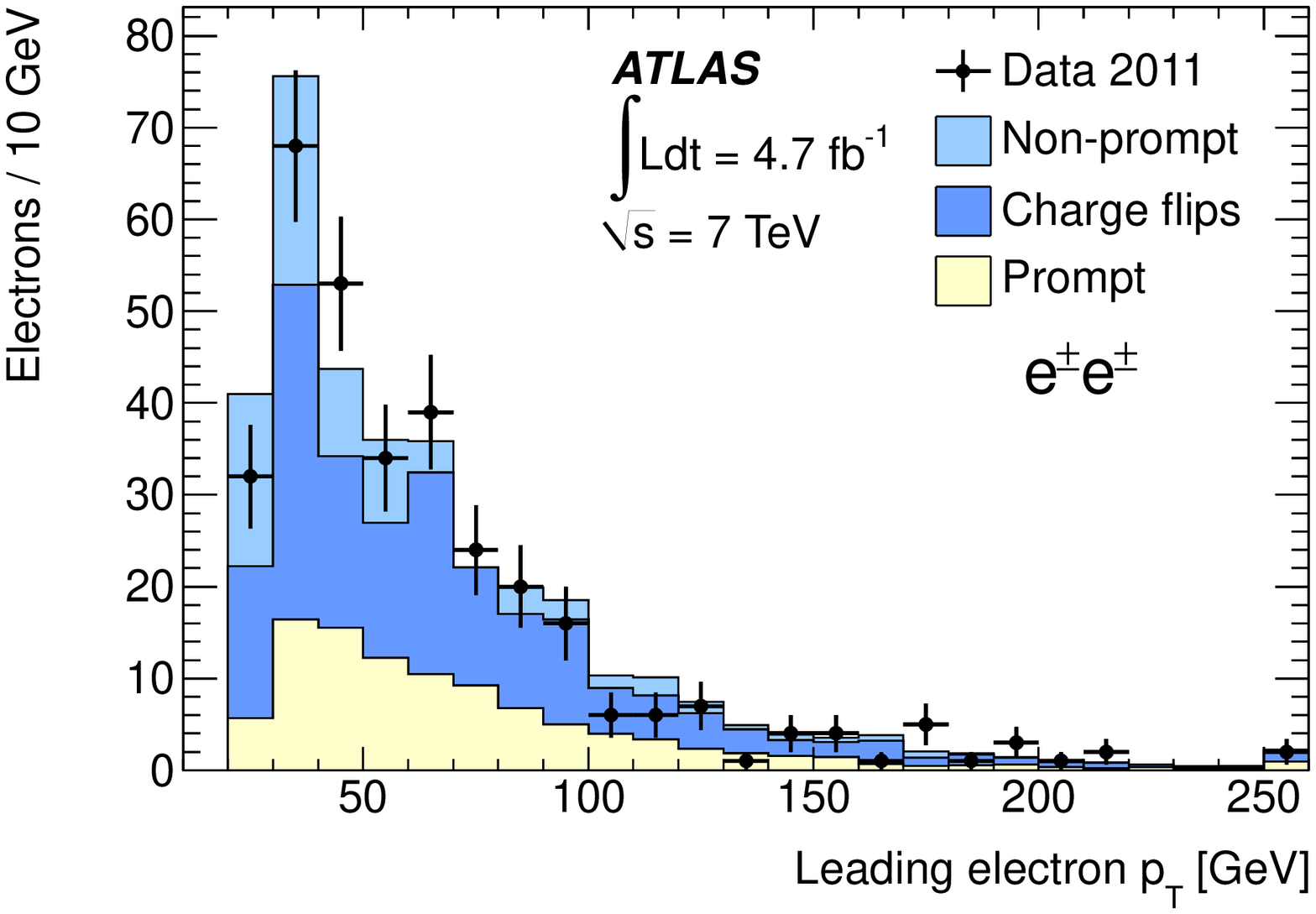}
}
\subfigure[] {
  \includegraphics[width=0.46\columnwidth]{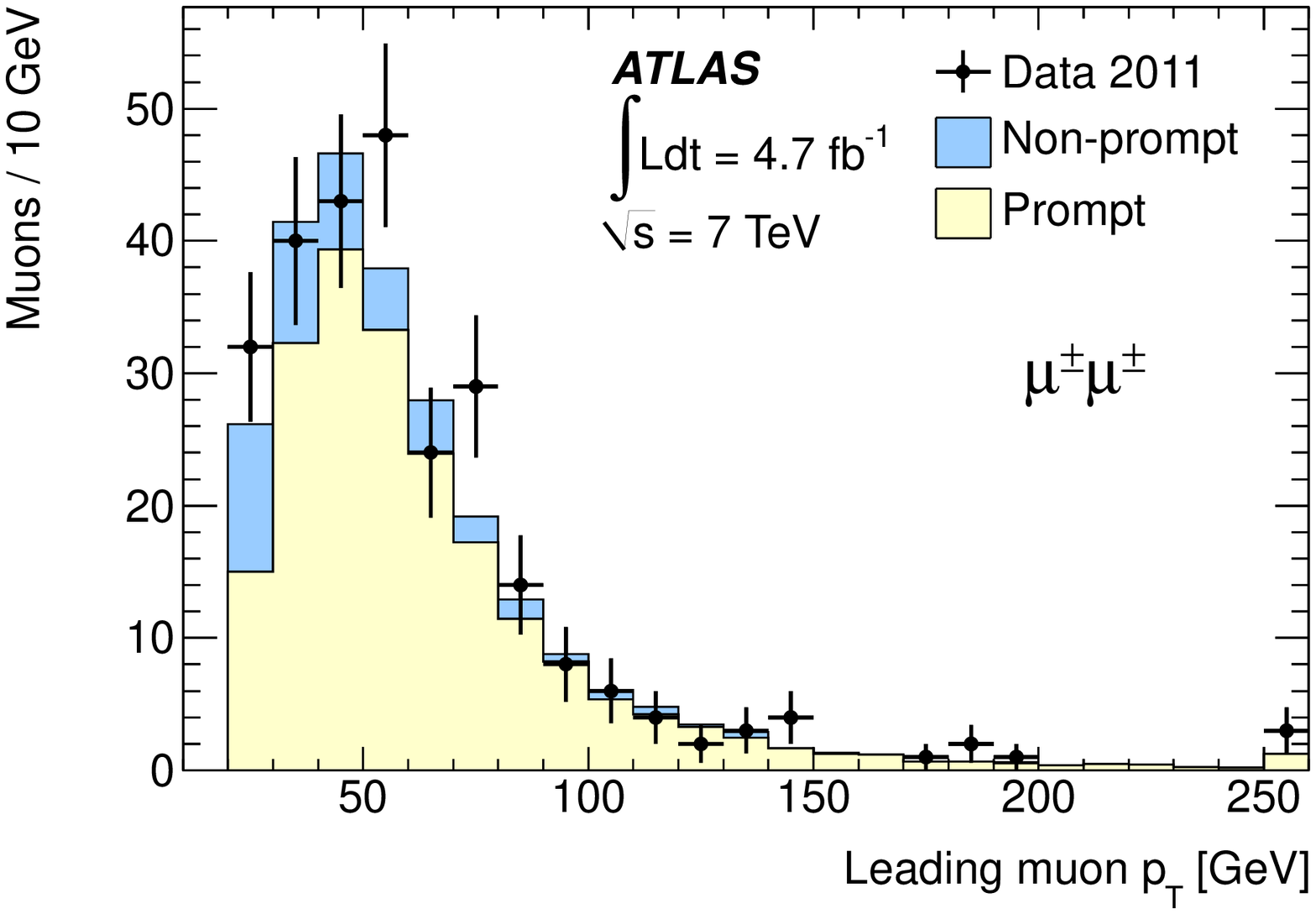}
}
\subfigure[] {
  \includegraphics[width=0.46\columnwidth]{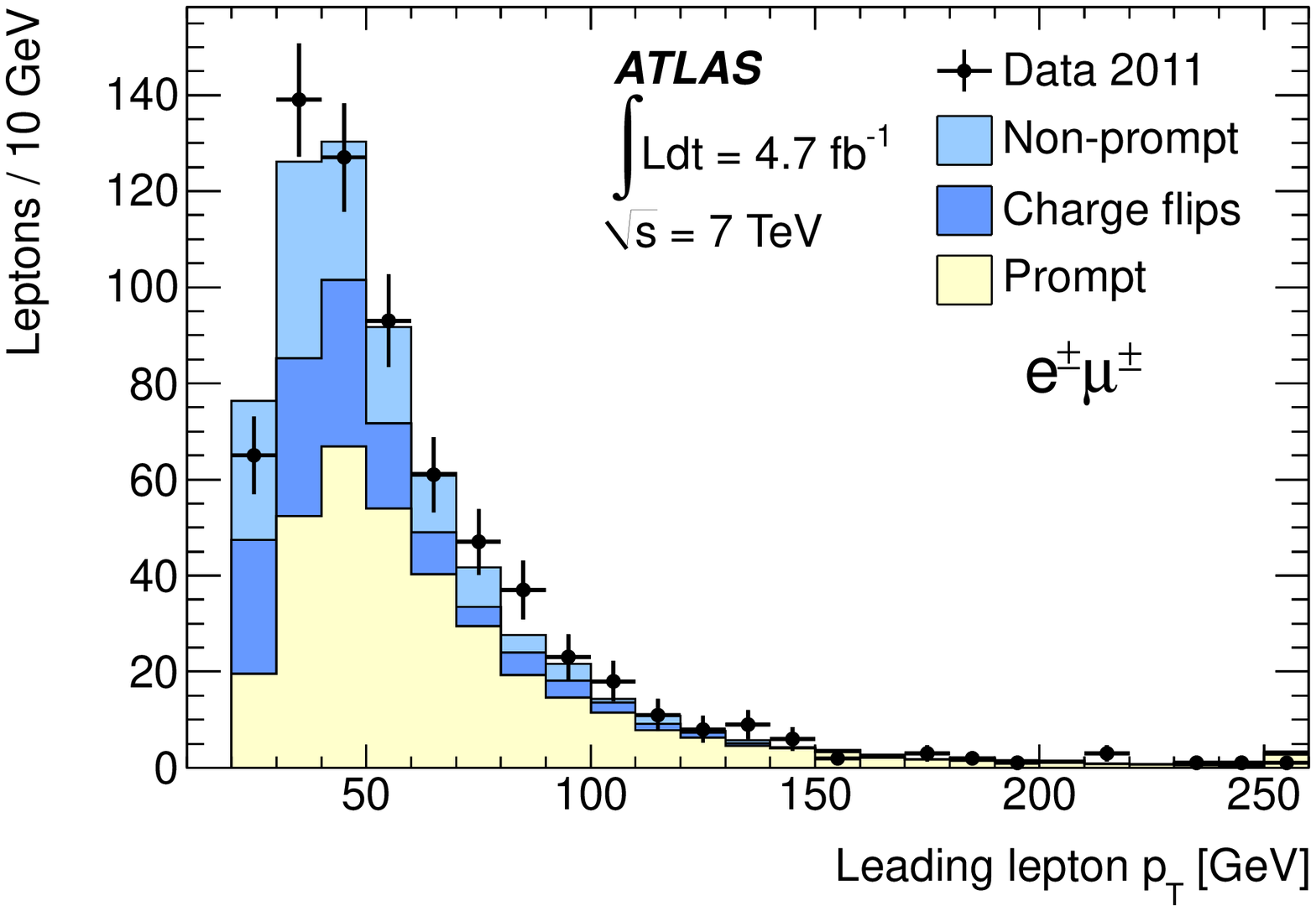}
}
\caption{\label{fig:dilep_pt_SS} Leading lepton $\pt$ distributions for (a) $\ee$, (b) $\mm$, and (c) $\emu$ pairs 
passing the full event selection. The data are shown as closed circles. The stacked histograms represent the backgrounds composed of pairs of prompt leptons from SM processes, pairs with at least one non-prompt lepton, and for the electron channels, backgrounds arising from charge misidentification and photon conversions. The last bin is an overflow bin.}
\end{center}
\end{figure}

\section{Upper limits on the cross section for prompt like-sign dilepton production}
\label{sec:limits}

Since the data are consistent with the background expectation, upper limits are placed on contributions due 
to processes from physics beyond the Standard Model. Limits are derived independently in each final state and mass bin shown in tables~\ref{tab:2iso_SS_both} and~\ref{tab:2iso_SS_sep}.  The 95\% confidence level (C.L.) upper limit on the number of events, $N_{95}$,
is determined using the CL$_s$ method~\cite{cls} with a test statistic based on the ratio between the likelihood of the signal plus background hypothesis and the likelihood of the background-only hypothesis.
The likelihoods follow a Poisson distribution for the total number of events in each signal region, and are calculated based on the number of expected and observed events.
Systematic uncertainties are incorporated into the likelihoods as nuisance parameters with Gaussian probability density functions.
For the inclusive $\ell^{\pm}\ell^{\pm}$ final states, the upper limit ranges 
from 168 pairs for $m(e\mu)>15$~GeV to 4.8 pairs for $m(\mu\mu)>400$~GeV. 

The limit on the number of lepton pairs can be translated to an upper limit on the cross section measured in a given region of phase space (referred to here as the fiducial region), $\sigma_{95}^{\mathrm{fid}}$, via
\begin{equation}
\label{eq:sigma}
\sigma_{95}^{\mathrm{fid}}= \frac{N_{95}}{\varepsilon_{\mathrm{fid}}\times \int {\cal L} dt},
\end{equation} 
where $\varepsilon_{\mathrm{fid}}$ is the efficiency for detecting events within the fiducial region
and $\int {\cal L} dt$ is the integrated luminosity of 4.7~fb$^{-1}$.

The fiducial region definition is based on MC generator information such that it is independent of the ATLAS detector environment.
The value of $\varepsilon_{\mathrm{fid}}$ generally depends on the new physics process, e.g. the number of leptons
in the final state passing the kinematic selection criteria or the number of jets 
that may affect the lepton isolation. In order to minimise this dependence,
the definition of the fiducial region is closely related to the analysis selection. 
In the MC generator, electrons or muons are selected as stable particles that originate from a $W$ or $Z$ boson, from a $\tau$ lepton,  
or from an exotic new particle (e.g. a $H^{\pm\pm}$ or a right-handed $W$). The generated
electrons and muons are required to satisfy the $\pt$, $\eta$, and isolation requirements
listed in table~\ref{tab:fid}, which mirror the selection requirements described in section~\ref{sec:evsel}.  
The generator-level track isolation, $p_{\rm T}^{{\rm cone} \Delta R_{\rm iso}}$, is defined as the scalar 
sum of all stable charged particles with $\pt>1$~GeV in a cone of size $\Delta R_{\rm iso}$ around the electron or muon, excluding the lepton itself.  
The requirement on isolation energy ($E_{\rm T}^{{\rm cone} \Delta R_{\rm iso}}$) placed in the electron selection is not emulated in the fiducial region definition because the measurement of isolation energy in the calorimeter has a poor resolution.
Pairs of like-sign leptons must have $m(\ell^{\pm}\ell^{\pm})>15$~GeV and for $\ee$, the mass range 70--110~GeV is additionally excluded. 

\begin{table}[htbp]
\caption{Summary of requirements on generated leptons in the fiducial region.  The definition of the isolation 
 variable, $p_{\rm T}^{{\rm cone} \Delta R_{\rm iso}}$, is given in the text.}
\begin{center}
\begin{tabular}{l|l|l}
 & Electron requirement & Muon requirement \\
\hline
Leading lepton $\pt$ & $\pt>25$~GeV & $\pt>20$~GeV \\
\hline
Sub-leading lepton $\pt$ & $\pt>20$~GeV & $\pt>20$~GeV \\
\hline
Lepton $\eta$ & $|\eta|<1.37$ or $1.52<|\eta|<2.47$ & $|\eta|<2.5$ \\
\hline
\multirow{2}{*}{Isolation} & \multirow{2}{*}{$p_{\rm T}^{\rm cone0.3}/\pt<0.1$} & $p_{\rm T}^{\rm cone0.4}/\pt<0.06$ and \\
 & & $p_{\rm T}^{\rm cone0.4}<4\mbox{ GeV}+0.02\times \pt$ \\
\hline
\end{tabular}
\end{center}
\label{tab:fid}
\end{table}

The fiducial efficiency, $\varepsilon_{\rm fid}$, is defined as the fraction of lepton pairs passing this selection
that also satisfy the experimental selection criteria described in section~\ref{sec:evsel}. 
It is determined for a variety of new physics models: $\hpp$ bosons with mass between 50~GeV and 1000~GeV~\cite{theory_hpphmm}, fourth
generation quarks decaying to $Wt$ with masses of 300--500~GeV, like-sign top-quark production via a contact interaction~\cite{AguilarSaavedra:2011zy}, 
and right-handed $W$ bosons decaying to $\ell^\pm\ell^\pm$ via a right-handed neutrino with $m(W_R)=$ 800--2500~GeV~\cite{theory_LRSM,theoryMNeu}. 
These models are chosen as they cover a broad range of jet multiplicities and lepton $\pt$ spectra.

For $m(\ell^{\pm}\ell^{\pm})>15$~GeV in the $\ee$ channel, the fiducial efficiencies range from 43\% for models with low-$\pt$ leptons to 65\% 
for models with high-$\pt$ leptons. The primary reason for this dependence is that the electron identification 
efficiency varies by about 15\% over the relevant $\pt$ range~\cite{epaper}.
The model dependence introduced by not emulating the calorimeter isolation in the definition of the fiducial region is $<1\%$.
For the $\emu$ channel, $\varepsilon_{\rm fid}$ ranges from 55\% to 70\%, and for the $\mm$ final state it varies between 59\% and 72\%.
For the higher dilepton mass thresholds the efficiencies are slightly larger than for the lower mass thresholds.  The efficiencies are also derived for $\ell^+\ell^+$ and $\ell^-\ell^-$ pairs separately and found to be independent of the charge.  For the same new physics models, the fraction of events satisfying the experimental selection originating from outside the fiducial region ranges from $<1\%$ to about 9\%, depending on the final state and the model considered.

To derive the upper limit on the fiducial cross section, the lowest efficiency values are taken for all mass thresholds,
i.e. 43\% for the $\ee$, 55\% for the $\emu$, and 59\% for the $\mm$ analysis. The 95\% C.L. upper limits 
on the cross section, $\sigma^{\rm fid}_{95}$, are calculated using equation~\ref{eq:sigma} and given in table~\ref{tab:lims_xsec}.
The limits range from 64~fb to 1.7~fb for the inclusive analysis depending on the mass cut and the final state, and are generally within $1\sigma$ of the expected limits.  Upper limits on the $\ell^+\ell^+$ and $\ell^-\ell^-$ cross sections are also derived, using the same efficiencies as for the inclusive limits.

\begin{table}[htbp]
\caption{Upper limits at 95\% C.L. on the fiducial cross section for $\ll$ pairs from non-SM physics.
The expected limits and their $1\sigma$ uncertainties are given, as well as the observed limits in data, 
for the $ee$, $e\mu$, and $\mu\mu$ final state inclusively and separated by charge.}
\begin{center}
\begin{tabular}{c||c|c||c|c||c|c}
\hline
 & \multicolumn{6}{c}{95\%  C.L. upper limit [fb]} \\
Mass range & expected & observed & expected & observed & expected & observed \\
 & \multicolumn{2}{c||}{$e^{\pm}e^{\pm}$} & \multicolumn{2}{c||}{$e^{\pm}\mu^{\pm}$} & \multicolumn{2}{c}{$\mu^{\pm}\mu^{\pm}$} \\
[+0.05in]
\hline
$m>15$~GeV
& $46^{+15}_{-12}$ & 42
& $56^{+23}_{-15}$ & 64
& $24.0^{+8.9}_{-6.0}$ & 29.8
 \\[+0.05in]
\hline
$m>100$~GeV
& $24.1^{+8.9}_{-6.2}$ & 23.4
& $23.0^{+9.1}_{-6.7}$ & 31.2
& $12.2^{+4.5}_{-3.0}$ & 15.0
 \\[+0.05in]
\hline
$m>200$~GeV
& $8.8^{+3.4}_{-2.1}$ & 7.5
& $8.4^{+3.4}_{-1.7}$ & 9.8
& $4.3^{+1.8}_{-1.1}$ & 6.7
 \\[+0.05in]
\hline
$m>300$~GeV
& $4.5^{+1.8}_{-1.3}$ & 3.9
& $4.1^{+1.8}_{-0.9}$ & 4.6
& $2.4^{+0.9}_{-0.7}$ & 2.6
 \\[+0.05in]
\hline
$m>400$~GeV
& $2.9^{+1.1}_{-0.8}$ & 2.4
& $3.0^{+1.0}_{-0.8}$ & 3.1
& $1.7^{+0.6}_{-0.5}$ & 1.7
 \\[+0.05in]
\hline\hline
 & \multicolumn{2}{c||}{$e^{+}e^{+}$} & \multicolumn{2}{c||}{$e^{+}\mu^{+}$} & \multicolumn{2}{c}{$\mu^{+}\mu^{+}$} \\
[+0.05in]
\hline
$m>15$~GeV
& $29.1^{+10.2}_{-8.6}$ & 22.8
& $34.9^{+12.2}_{-8.6}$ & 34.1
& $15.0^{+6.1}_{-3.3}$ & 15.2
 \\[+0.05in]
\hline
$m>100$~GeV
& $16.1^{+5.9}_{-4.3}$ & 12.0
& $15.4^{+5.9}_{-4.1}$ & 18.0
& $8.4^{+3.2}_{-2.4}$ & 7.9
 \\[+0.05in]
\hline
$m>200$~GeV
& $7.0^{+2.9}_{-2.2}$ & 6.1
& $6.6^{+3.5}_{-1.8}$ & 8.8
& $3.5^{+1.6}_{-0.7}$ & 4.3
 \\[+0.05in]
\hline
$m>300$~GeV
& $3.7^{+1.4}_{-1.0}$ & 2.9
& $3.2^{+1.2}_{-0.9}$ & 3.2
& $2.0^{+0.8}_{-0.5}$ & 2.1
 \\[+0.05in]
\hline
$m>400$~GeV
& $2.3^{+1.1}_{-0.6}$ & 1.7
& $2.4^{+0.9}_{-0.6}$ & 2.5
& $1.5^{+0.6}_{-0.3}$ & 1.8
 \\[+0.05in]
\hline\hline
 & \multicolumn{2}{c||}{$e^{-}e^{-}$} & \multicolumn{2}{c||}{$e^{-}\mu^{-}$} & \multicolumn{2}{c}{$\mu^{-}\mu^{-}$} \\
[+0.05in]
\hline
$m>15$~GeV
& $23.2^{+8.6}_{-5.8}$ & 25.7
& $26.2^{+10.6}_{-7.6}$ & 34.4
& $12.1^{+4.5}_{-3.5}$ & 18.5
 \\[+0.05in]
\hline
$m>100$~GeV
& $12.0^{+5.3}_{-2.8}$ & 18.7
& $11.5^{+4.2}_{-3.5}$ & 16.9
& $6.0^{+2.3}_{-1.9}$ & 10.1
 \\[+0.05in]
\hline
$m>200$~GeV
& $4.9^{+1.9}_{-1.2}$ & 4.0
& $4.6^{+2.1}_{-1.2}$ & 4.5
& $2.7^{+1.1}_{-0.7}$ & 4.4
 \\[+0.05in]
\hline
$m>300$~GeV
& $2.9^{+1.0}_{-0.6}$ & 2.7
& $2.7^{+1.1}_{-0.6}$ & 3.5
& $1.5^{+0.8}_{-0.3}$ & 1.7
 \\[+0.05in]
\hline
$m>400$~GeV
& $1.8^{+0.8}_{-0.4}$ & 2.3
& $2.3^{+0.8}_{-0.5}$ & 2.5
& $1.2^{+0.4}_{-0.0}$ & 1.2
 \\[+0.05in]
\hline
\end{tabular}
\end{center}
\label{tab:lims_xsec}
\end{table}

\section{Conclusions}
\label{sec:conclusions}

A search for anomalous production of like-sign lepton pairs has been presented using 4.7~fb$^{-1}$ 
of $pp$ collision data at $\sqrt{s}=7$~TeV recorded by the ATLAS experiment at the LHC. 
The data are found to agree with the background expectation in $\ee$, $\emu$, and $\mm$ final states 
both in overall rate and in the kinematic distributions. The data are used to constrain new physics contributions 
to like-sign lepton pairs within a fiducial region of two isolated leptons with large transverse momentum within the pseudorapidity range of the tracking system ($|\eta|<2.5$). The 95\% confidence level upper limits on the cross section of new physics processes within this fiducial region range
between 1.7~fb and 64~fb for $\ell^{\pm}\ell^{\pm}$ pairs depending on the dilepton invariant mass and flavour combination.

\section{Acknowledgements}

We thank CERN for the very successful operation of the LHC, as well as the
support staff from our institutions without whom ATLAS could not be
operated efficiently.

We acknowledge the support of ANPCyT, Argentina; YerPhI, Armenia; ARC,
Australia; BMWF and FWF, Austria; ANAS, Azerbaijan; SSTC, Belarus; CNPq and FAPESP,
Brazil; NSERC, NRC and CFI, Canada; CERN; CONICYT, Chile; CAS, MOST and NSFC,
China; COLCIENCIAS, Colombia; MSMT CR, MPO CR and VSC CR, Czech Republic;
DNRF, DNSRC and Lundbeck Foundation, Denmark; EPLANET and ERC, European Union;
IN2P3-CNRS, CEA-DSM/IRFU, France; GNSF, Georgia; BMBF, DFG, HGF, MPG and AvH
Foundation, Germany; GSRT, Greece; ISF, MINERVA, GIF, DIP and Benoziyo Center,
Israel; INFN, Italy; MEXT and JSPS, Japan; CNRST, Morocco; FOM and NWO,
Netherlands; BRF and RCN, Norway; MNiSW, Poland; GRICES and FCT, Portugal; MERYS
(MECTS), Romania; MES of Russia and ROSATOM, Russian Federation; JINR; MSTD,
Serbia; MSSR, Slovakia; ARRS and MVZT, Slovenia; DST/NRF, South Africa;
MICINN, Spain; SRC and Wallenberg Foundation, Sweden; SER, SNSF and Cantons of
Bern and Geneva, Switzerland; NSC, Taiwan; TAEK, Turkey; STFC, the Royal
Society and Leverhulme Trust, United Kingdom; DOE and NSF, United States of
America.

The crucial computing support from all WLCG partners is acknowledged
gratefully, in particular from CERN and the ATLAS Tier-1 facilities at
TRIUMF (Canada), NDGF (Denmark, Norway, Sweden), CC-IN2P3 (France),
KIT/GridKA (Germany), INFN-CNAF (Italy), NL-T1 (Netherlands), PIC (Spain),
ASGC (Taiwan), RAL (UK) and BNL (USA) and in the Tier-2 facilities
worldwide.

\bibliographystyle{atlasnote}
\providecommand{\href}[2]{#2}\begingroup\raggedright\endgroup

\onecolumn
\clearpage
\input{atlas_authlist}

\end{document}

%% file: atlas_authlist.tex
\begin{flushleft}
{\Large The ATLAS Collaboration}

\bigskip

G.~Aad$^{\rm 48}$,
T.~Abajyan$^{\rm 21}$,
B.~Abbott$^{\rm 111}$,
J.~Abdallah$^{\rm 12}$,
S.~Abdel~Khalek$^{\rm 115}$,
A.A.~Abdelalim$^{\rm 49}$,
O.~Abdinov$^{\rm 11}$,
R.~Aben$^{\rm 105}$,
B.~Abi$^{\rm 112}$,
M.~Abolins$^{\rm 88}$
% ...
\bigskip

% ...
$^{11}$ Institute of Physics, Azerbaijan Academy of Sciences, Baku, Azerbaijan\\
$^{12}$ Institut de F\'{i}sica d'Altes Energies and Departament de F\'{i}sica de la Universitat Aut\`{o}noma de Barcelona and ICREA, Barcelona, Spain\\
% ...
$^{21}$ Physikalisches Institut, University of Bonn, Bonn, Germany\\
% ...
$^{48}$ Fakult\"{a}t f\"{u}r Mathematik und Physik, Albert-Ludwigs-Universit\"{a}t, Freiburg, Germany\\
$^{49}$ Section de Physique, Universit\'{e} de Gen\`{e}ve, Geneva, Switzerland\\
% ...
$^{88}$ Department of Physics and Astronomy, Michigan State University, East Lansing MI, United States of America\\
% ...
$^{105}$ Nikhef National Institute for Subatomic Physics and University of Amsterdam, Amsterdam, Netherlands\\
% ...
$^{111}$ Homer L. Dodge Department of Physics and Astronomy, University of Oklahoma, Norman OK, United States of America\\
$^{112}$ Department of Physics, Oklahoma State University, Stillwater OK, United States of America\\
% ...
$^{115}$ LAL, Universit\'{e} Paris-Sud and CNRS/IN2P3, Orsay, France\\
% ...
\end{flushleft}
